\documentclass[aps,showpacs,prb14b,floatfix,twocolumn,citeautoscript]{revtex4-1}
\usepackage{graphicx}
\usepackage{bm,amsmath,amssymb,mathrsfs,dcolumn}
\usepackage{subfigure}
\usepackage{units}
\usepackage{hyperref}
\usepackage[usenames]{color}
\hypersetup{pdfborder=0 0 0,colorlinks=true,citecolor=blue,linkcolor=blue}
\input epsf

\newcommand{\BN}{\textit{h}-BN}

\begin{document}

\title{Large potential steps at weakly interacting metal-insulator interfaces}

\author{Menno Bokdam}
\altaffiliation{Current address: Faculty of Physics, University of Vienna, 
Computational Materials Physics, Sensengasse 8/12, 1090 Vienna, Austria.}
\author{Geert Brocks}
\author{Paul J. Kelly}
\affiliation{Faculty of Science and Technology and MESA$^{+}$ Institute for
Nanotechnology, University of Twente, P.O. Box 217, 7500 AE Enschede, The
Netherlands}

\begin{abstract}
Potential steps exceeding 1 eV are regularly formed at metal$|$insulator 
interfaces, even when the interaction between the materials at the interface is 
weak physisorption. From first-principles calculations on metal$|$\BN{} 
interfaces we show that these potential steps are only indirectly sensitive to 
the interface bonding through the dependence of the binding energy curves on the 
van der Waals interaction. Exchange repulsion forms the main contribution to the 
interface potential step in the weakly interacting regime, which we show  
 with a simple model based upon a symmetrized product of metal and 
\BN{} wave functions. In the strongly interacting regime, the interface 
potential step is reduced by chemical bonding.
\end{abstract}
\date{\today}
\maketitle

{\color{red}\it Introduction.}---The potential step that is formed at the 
interface between a metal and a semiconductor or insulator is an essential 
physical parameter determining device performance \cite{Tung:apr14}. A 
satisfactory understanding of the factors influencing the step is complicated by 
its extreme sensitivity to interface structure and disorder \cite{Tung:apr14}. 
Metal contacts with the layered van der Waals (vdW) structures 
that are the subject of much current study 
\cite{Geim:nat13,*Popov:prl12,*Chen:nanol13,*Gong:nanol14,*Kang:prx14}, in 
particular hexagonal boron nitride ($h$-BN) \cite{Dean:natn10}, form an ideal 
model system to study such metal contacts. Because $h$-BN is chemically 
unreactive, it can form essentially defect-free interfaces with metals making 
possible a particularly clean confrontation of theory with experiment. Potential 
steps involving $h$-BN are already very interesting in their own right 
\cite{Geim:nat13,Dean:natn10}.

\begin{figure} [b]
\includegraphics[width=8.5cm]{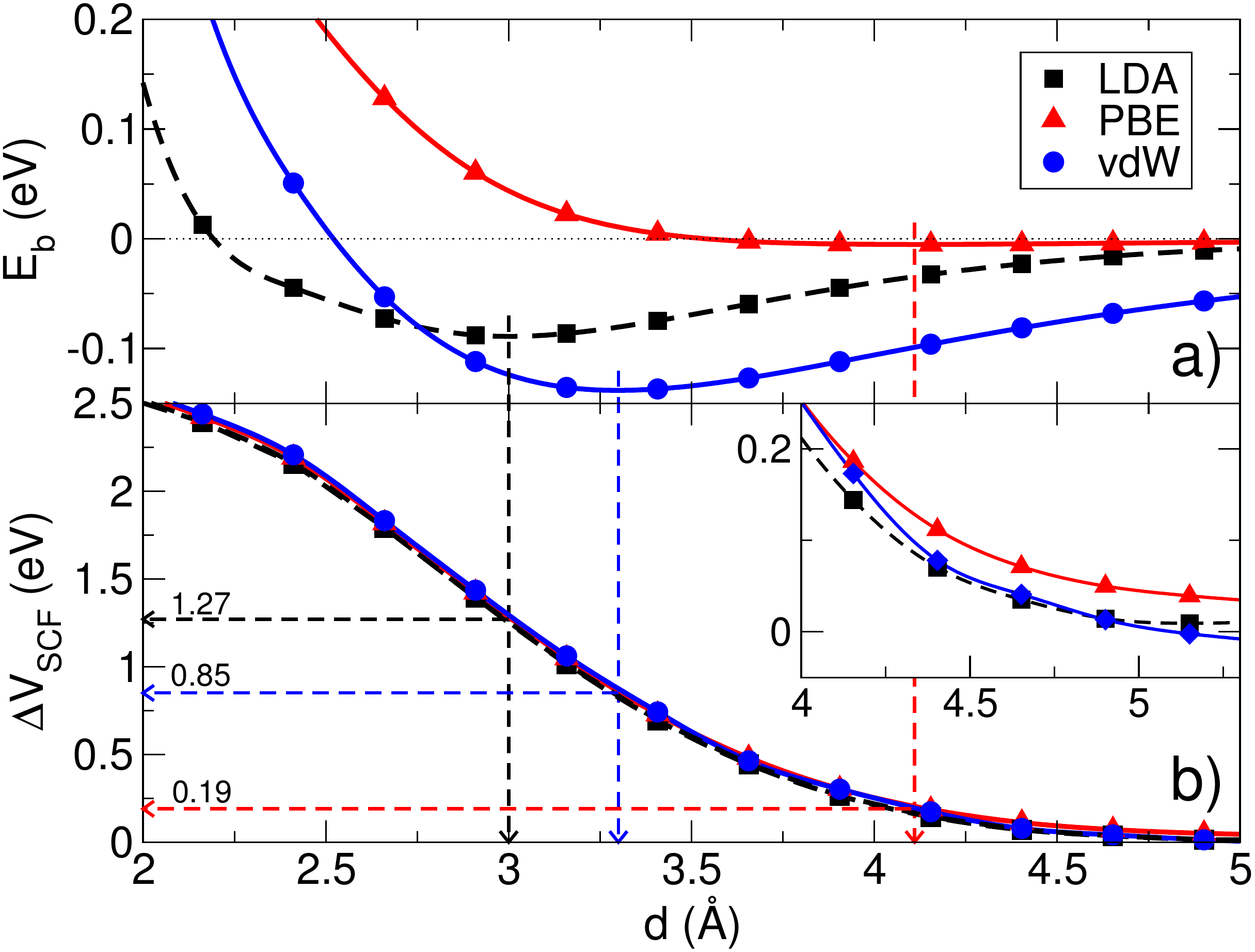}
\caption{(color online) (a) Binding energy curves ${\rm E_b}(d)$ in eV/BN for a monolayer of \BN{} on Cu(111) calculated with LDA (black squares), GGA-PBE (red triangles), and optB88-vdW-DF (blue circles) functionals. The vertical dashed lines indicate the minima. (b) The corresponding interface potential steps $\Delta V_{\rm SCF}(d)$ calculated with the three functionals.}
\label{fig1}
\end{figure}

The formation of an interface between metals and 2D materials such as graphene 
or \BN{} leads to a dipole layer and a potential step at the interface 
\cite{Nagashima:ss96,Preobrajenski:ss05,Leuenberger:prb11,Joshi:nanol12}. 
Naively, one might expect the interface dipole and potential step to be small if 
the interaction between the two materials is weak. It is then puzzling to find 
that physisorption of graphene or \BN{} on metal substrates, with adsorption 
energies as small as $\sim 0.05$~eV/atom \cite{Olsen:prl11,Stradi:prl11}, leads 
to substantial potential steps $\Delta V$ of $\sim1$~eV 
\cite{Giovannetti:prl08,*Khomyakov:prb09,Bokdam:nanol11,*Bokdam:prb13}, see 
Figure \ref{fig1}. A possible explanation for a large potential step is direct 
transfer of electrons across the interface, which occurs on equilibrating the 
chemical potential between two conductors. This happens at metal$|$graphene 
interfaces, for instance, and results in doping of graphene; the corresponding 
contribution to the potential step is $\Delta_{\rm tr}$ 
\cite{Giovannetti:prl08}. 
For graphene there is an additional, large contribution to the total potential step arising from the direct (physisorption) interaction at the interface. That 
contribution, called $\Delta_{c}$ in Ref.~\onlinecite{Khomyakov:prb09}, was 
found to depend roughly exponentially on the graphene-metal distance, 
underlining its local, interface character. Thus, for graphene, 
$ \Delta V= \Delta_{\rm tr} + \Delta_{\rm c}$. Similar 
terms were identified in Ref.~\onlinecite{Gebhardt:prb12}.
Direct charge transfer cannot occur across a 
metal$|$\BN{} interface because \BN{} is a wide band gap insulator. Yet even 
here large potential steps are found 
\cite{Joshi:nanol12,Nagashima:prl95,Bokdam:prb13,Schulz:prb14}. The 
absence of direct charge transfer makes it possible to study the potential step 
arising from just the interface interaction.

In this paper, we use first-principles calculations to explore the 
origin of the interface dipole and potential step at metal$|$insulator 
interfaces. We focus on metal$|$\BN{} as an archetypal interface, selecting in 
particular those cases where the bonding interaction at the interface is weak. 
Surprisingly, the interface potential step does not depend on the 
exchange-correlation functional used to calculate it, even though the binding 
energy curve is quite sensitive to that functional as shown by Fig.~\ref{fig1}. 
This points to a more general origin of the potential step. From a transparent 
model based upon a symmetrized product of fragment states, we show that exchange 
repulsion at the interface between the metal and the insulator is the main 
source of the potential step. The van der Waals bonding between the two 
materials gives a much smaller contribution.  

Interactions between closed-shell molecules or ions give rise to dipoles with an 
exponential separation dependence \cite{Brocks:jcp84,*Brocks:molphys85}. This 
behavior is ascribed to Pauli repulsion which pushes electrons out of the 
overlap region between molecules and results in a distortion of the electron 
distribution. The dipoles that are formed when inert atoms or molecules adsorb 
on metal surfaces are also attributed to Pauli repulsion. In this context, the 
occurrence of such dipoles is called the push-back or pillow effect 
\cite{Dresser:ss74,*Bagus:prl02,*Vazquez:jcp07,*Bagus:prl08,*Vitali:natm10,
*Rusu:prb10,Gebhardt:prb12}. In the following, we demonstrate the effect of the 
Pauli repulsion explicitly by calculating its contribution to the potential 
steps at weakly interacting metal$|$h-BN interfaces.

{\color{red}\it DFT calculations.}---The potential step at an A$|$B 
metal$|$insulator interface is obtained from self-consistent calculations as the 
difference between the work functions $W$ of the clean metal surface, $W_{\rm 
A}$, and of the combined system; $\Delta V_{\rm SCF} = W_{\rm A} - W_{\rm 
A|B}$ \cite{Neugebauer:prb92}. Here we adsorb a \BN{} monolayer on 
close-packed (111) metal surfaces. We consider commensurable interfaces, 
accommodating lattice mismatch by adapting the in-plane lattice constant of the 
metals to that of \BN{} so that the strain remains $< 5$\%, as in 
Refs.~\onlinecite{Khomyakov:prb09} and \onlinecite{Bokdam:prb13}. Changing the 
lattice constant of a metal by a few percent changes its electronic properties 
only mildly, whereas adapting the lattice constant of 
\BN{} is a much larger perturbation. Weakly interacting 
metal$|h$-BN interfaces are found to exhibit in-plane moir\'e patterns with 
large periods. Calculations for such in-plane superstructures  are 
computationally very demanding and are not crucial for the present study. A more 
detailed discussion of the effects of incommensurability can be found in 
Refs.~\onlinecite{GomezDiaz:tca13,Bokdam:prb14b}. 

For the DFT calculations we use the Vienna Ab-initio Simulation Package ({\sc 
vasp}) \cite{Kresse:prb93,*Kresse:prb96,*Kresse:prb99}, and follow 
Ref.~\onlinecite{Bokdam:prb13} concerning interface structures and choice of 
computational parameters. We 
consider three different functionals: the local density approximation (LDA) 
\cite{Perdew:prb81}, the PBE generalized gradient approximation (GGA) 
\cite{Perdew:prl96}, and a van der Waals density functional 
\cite{Dion:prl04,Thonhauser:prb07,Klimes:prb11}. Though it generally 
overestimates chemical interactions, the LDA gives a reasonable description of 
the binding energy and equilibrium separation of metal$|$\BN{} 
interfaces 
\cite{Bokdam:prb14b}. GGA often gives a good description of 
chemisorption but fails to capture physisorption. Local or semi-local 
functionals lack vdW interactions which play an important role in physisorption 
\cite{Ruiz:prl12} and in the bonding of layered materials \cite{Bjorkman:prl12}. 
These interactions are modeled in non-local vdW functionals. Here we use the 
optB88-vdW-DF functional \cite{Klimes:prb11}, which has been shown 
to give a good description of graphene on Ni \cite{Mittendorfer:prb11}.

Figure \ref{fig1}(a) shows the binding energy curves of \BN{} on Cu(111) for the 
three functionals. GGA gives essentially no bonding, with an adsorption energy 
$E_{\rm b} = -1$ meV/BN at an equilibrium separation $d_{\rm eq} = 4.1$ \AA; LDA 
gives a reasonable bonding with $E_{\rm b} = -87$ meV/BN at $d_{\rm eq} = 3.0$ 
\AA; optB88-vdW-DF gives $E_{\rm b} = -140$ meV/BN at $d_{\rm eq} = 3.3$ 
\AA, underlining the importance of vdW interactions 
\cite{Reguzzoni:prb12,Berland:prb13}. Whereas the binding energy curves 
evidently depend sensitively on the (type of) functional used 
\footnote{Experiments concur that the interaction is weak but the 
binding separation has not been measured 
directly \cite{Preobrajenski:ss05,Joshi:nanol12,Roth:nanol13}; the value 
reported in Ref.~\cite{Joshi:nanol12} results 
from interpreting STM measurements with the help of DFT calculations.}, the 
interface potential step is remarkably insensitive. This is clearly demonstrated 
in Fig.~\ref{fig1}(b) where the potential step at the Cu(111)$|$\BN{} interface, 
$\Delta V_{\rm SCF}$, is shown as a function of the separation $d$ between the 
Cu(111) surface and the \BN{} plane. The curves for the three functionals are 
within 0.05 eV of one another. 

\begin{figure}
\includegraphics[width=8.5cm]{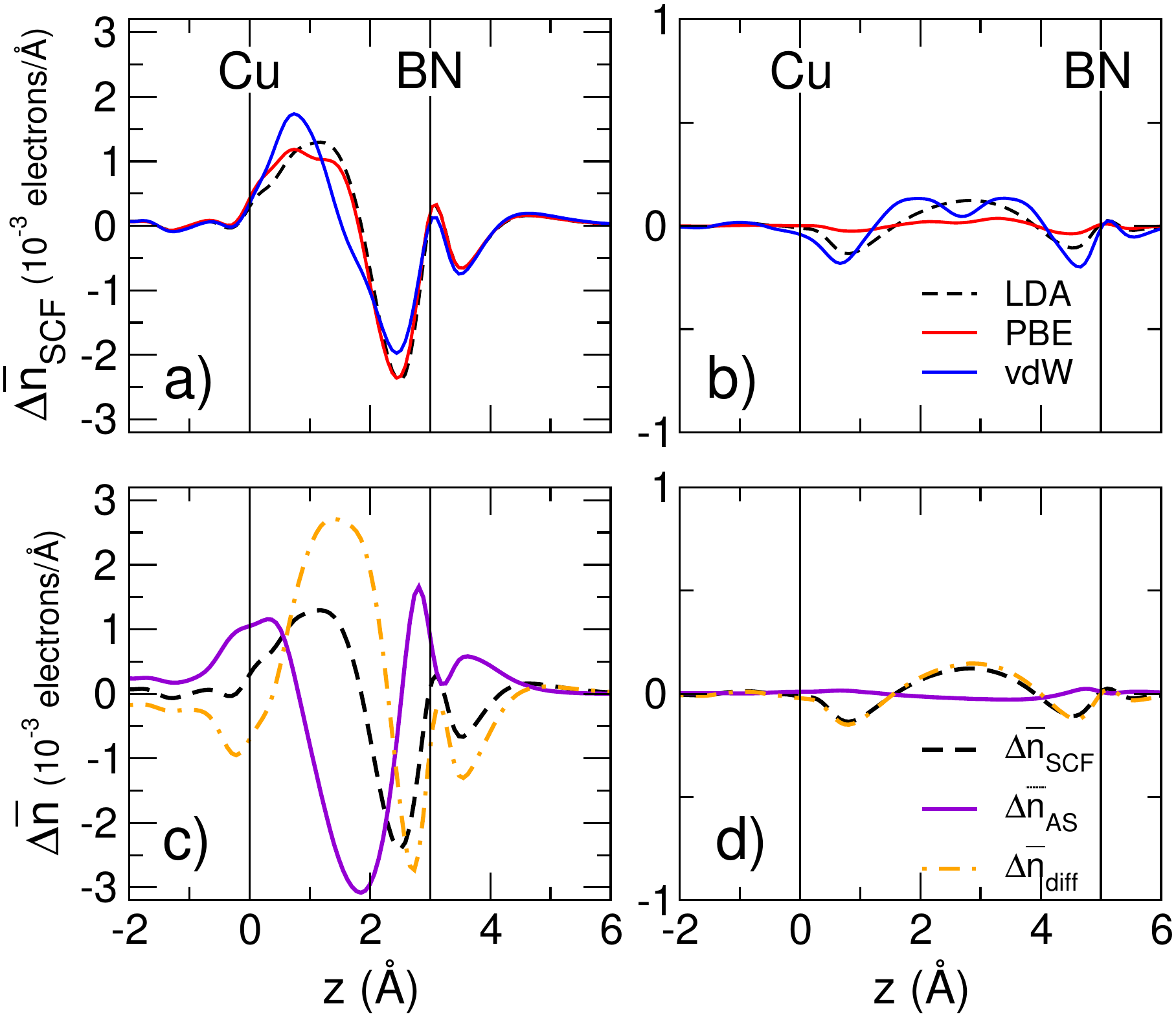}
\caption{(color online)
The plane-averaged electron displacements from self-consistent calculations 
$\Delta \overline{n}_{\rm SCF}(z)$ at (a) $d=3$~\AA\ and (b) $d=5$ \AA.
(c) Comparison of $\Delta\overline{n}_{\rm SCF}$ (LDA: black 
dashed) with $\Delta\overline{n}_{\rm AS}$ of the corresponding AS state 
calculated 
using Eq.~(\ref{eq:10}) (violet); and their difference 
$\Delta\overline{n}_{\rm diff}$ 
(orange dash-dotted) referred to as the bonding contribution, for a 
Cu(111)--\BN{} separation of 3 \AA\ and (d) 5 \AA.
}
\label{fig2}
\end{figure}

The potential step is proportional to the interface dipole which can be derived 
from the electron displacement $\Delta n_{{\rm SCF}} = n_{\rm A|B}-n_{\rm 
A}-n_{\rm B}$, where $n_{\rm A|B},n_{\rm A}$, and $n_{\rm B}$ are, respectively, 
the electron densities of the metal$|$\BN{} system, the isolated metal, and the 
\BN{} monolayer. The insensitivity of the potential step to the functional 
suggests a similar insensitivity of $\Delta n_{{\rm SCF}}$, which is confirmed 
by Figs.~\ref{fig2}(a) and (b). At $d=3$ \AA, i.e., close to the 
(experimental) equilibrium separation proposed in 
Ref.~\onlinecite{Joshi:nanol12}, $\Delta V_{{\rm SCF}} \approx 1$ eV, and the 
plane-averaged electron displacement, $\Delta \overline{n}_{{\rm SCF}}(z)$, is 
very similar for all three functionals. At a larger separation, $d=5$ \AA, 
$\Delta \overline{n}_{{\rm SCF}}$ for the vdW-DF functional shows an 
accumulation of electrons between the top metal plane and the \BN{} sheet, and a 
depletion closer to these. 

Such a pattern is also observed in an $(\mathrm{Ar})_2$ vdW complex, suggesting that this is typical for vdW interactions \cite{Thonhauser:prb07}. Indeed this pattern is absent from the PBE $\Delta \overline{n}_{{\rm SCF}}$ at $d=5$ \AA. Interestingly,  $\Delta \overline{n}_{{\rm SCF}}$ calculated with the LDA is quite similar to that found for the vdW-DF. A local functional cannot represent vdW interactions properly, hence the substantial differences between the LDA and the vdW-DF binding energy curves. Nevertheless, it is remarkable that both the LDA and vdW-DF give very similar electron distributions. At $d=5$ \AA, $\Delta V_{{\rm SCF}} < 0.05$ eV, demonstrating that vdW interactions as such do not give rise to large interface dipoles. 

{\color{red}\it Model.}---The insensitivity of the dipole to the functional used suggests a model that does not rely heavily upon the specific functional. An approximation to the ground state of an A$|$B interface should be a fermion state. Starting from two well-separated systems A and B, a simple fermion state is the antisymmetrized product
\begin{equation}
\left|\Psi\right\rangle =\hat{A}\hat{B}\left|0\right\rangle,\label{eq:1}
\end{equation}
where 
\begin{equation}
\hat{X}=\prod_{\mathbf{k}n \in {\rm occ}}\hat{c}_{X,\mathbf{k}n}^{\dagger} \;\;\; ; \; X=A,B,\label{eq:2}
\end{equation}
with $\left|0\right\rangle $ the vacuum, $\mathbf{k}n$ the Bloch vector and band index. The fermion operator $\hat{c}_{X,\mathbf{k}n}^{\dagger}$ creates an electron in the orbital $\left|\phi_{\mathbf{k}n}^{X}\right\rangle$, and the product is over all occupied states. The state $\left|\Psi\right\rangle$ incorporates the exchange of electrons among any of the occupied orbitals of A and B. We take it to define the Pauli exchange interaction between systems A and B, and refer to this state as the anti-symmetrized (AS) product state.

If the orbitals on A and B overlap at the interface between the two systems, they are in general not orthogonal. The technical difficulties of calculating expectation values with non-orthogonal orbitals can be circumvented. Define a linear transformation, 
${{\hat{c}}'^{\dagger}}_{\beta} =\sum_{\alpha}\hat{c}_{\alpha}^{\dagger}T_{\alpha\beta}$, 
where $\alpha$ or $\beta$ is a short-hand notation for the combined index $(X,\mathbf{k}n)$ that runs over all occupied states of both systems A and B. The same transformation defines new orbitals, $|\phi'_{\beta}\rangle=\sum_{\beta}|\phi_{\alpha}\rangle T_{\alpha\beta}$. The state $\left|\Psi\right\rangle$ is invariant under such a transformation, apart from a multiplicative factor, $\left|\Psi'\right\rangle = \det(T)\left|\Psi\right\rangle$, which follows directly from its definition, Eqs. (\ref{eq:1}) and (\ref{eq:2}). Orthogonalizing the orbitals $\langle \phi'_{\alpha}|\phi'_{\beta}\rangle =\delta_{\alpha\beta}=\sum_{\gamma,\zeta}T_{\gamma\alpha}^{*}S_{\gamma\zeta}T_{\zeta\beta}$, where 
$S_{\gamma\zeta}=\left\langle \phi{}_{\gamma}|\phi{}_{\zeta}\right\rangle$ is the overlap matrix of the original orbitals, defines a transformation that reads in matrix form $\mathbf{I}=\mathbf{T}^{\dagger}\mathbf{ST}$, or $\mathbf{TT}^{\dagger}=\mathbf{S}^{-1}$. The expectation value with respect to $\left|\Psi\right\rangle $ of any operator can then be calculated using the standard expressions for orthogonal orbitals. For instance, for any single-particle operator one obtains 
$
\sum_{\alpha}\left\langle \phi'_{\alpha}\left|\hat{o}\right|\phi'_{\alpha}\right\rangle =\sum_{\alpha,\beta}\left\langle \phi{}_{\alpha}\left|\hat{o}\right|\phi{}_{\beta}\right\rangle S_{\beta\alpha}^{-1}.
$

The density operator $\hat{n}(\mathbf{r})=\left|\mathbf{r}\right\rangle \left\langle \mathbf{r}\right|$ is an example of a single-particle operator,
whose expectation value is the electron density
\begin{equation}
n_{\rm AS}(\mathbf{r})=\sum_{\alpha,\beta}\phi_{\alpha}^{*}(\mathbf{r})\phi{}_{\beta}(\mathbf{r})S_{\beta\alpha}^{-1},\label{eq:9}
\end{equation}
with $\phi_{\alpha}(\mathbf{r})\equiv\left\langle \mathbf{r}|\phi_{\alpha}\right\rangle $. We define the electron displacement $\Delta n_{\rm AS}(\mathbf{r})$ as the change in the electron density of
the combined AB system with respect to the sum of the electron densities of the two separate systems, A and B,  
\begin{equation}
\Delta n_{\rm AS}(\mathbf{r})=\sum_{\alpha,\beta}\phi_{\alpha}^{*}(\mathbf{r})\phi{}_{\beta}(\mathbf{r})\left(S_{\beta\alpha}^{-1}-\delta_{\beta\alpha}\right).\label{eq:10}
\end{equation}
The double sum is over all occupied orbitals $(X,\mathbf{k}n)$. The overlap matrix and its inverse are block-diagonal in $\mathbf{k}$, but not in the system and band indices $X,n$.

By construction, $\int\Delta n_{\rm AS}(\mathbf{r})d^{3}r=0$ (integrated over all space). 
If the overlap between the A and B subsystems is confined to an interface, then 
$\Delta n_{\rm AS}(\mathbf{r})\rightarrow 0$ away from the interface. Solving the Poisson 
equation with $\Delta n_{\rm AS}(\mathbf{r})$ as source then yields a step in the 
potential (energy) across the interface. Averaging $\Delta n_{\rm AS}(\mathbf{r})$ over 
planes yields $\Delta\overline{n}_{\rm AS}(z)$  in terms of which the step $\Delta 
V_{\rm AS}=\frac{e^{2}}{\epsilon_{0}}\int_{-\infty}^{\infty}z 
\Delta\overline{n}_{\rm AS}(z)dz$ can be defined, with $z$ the direction normal to the 
interface.

\begin{figure}[b]
\includegraphics[width=8.5cm]{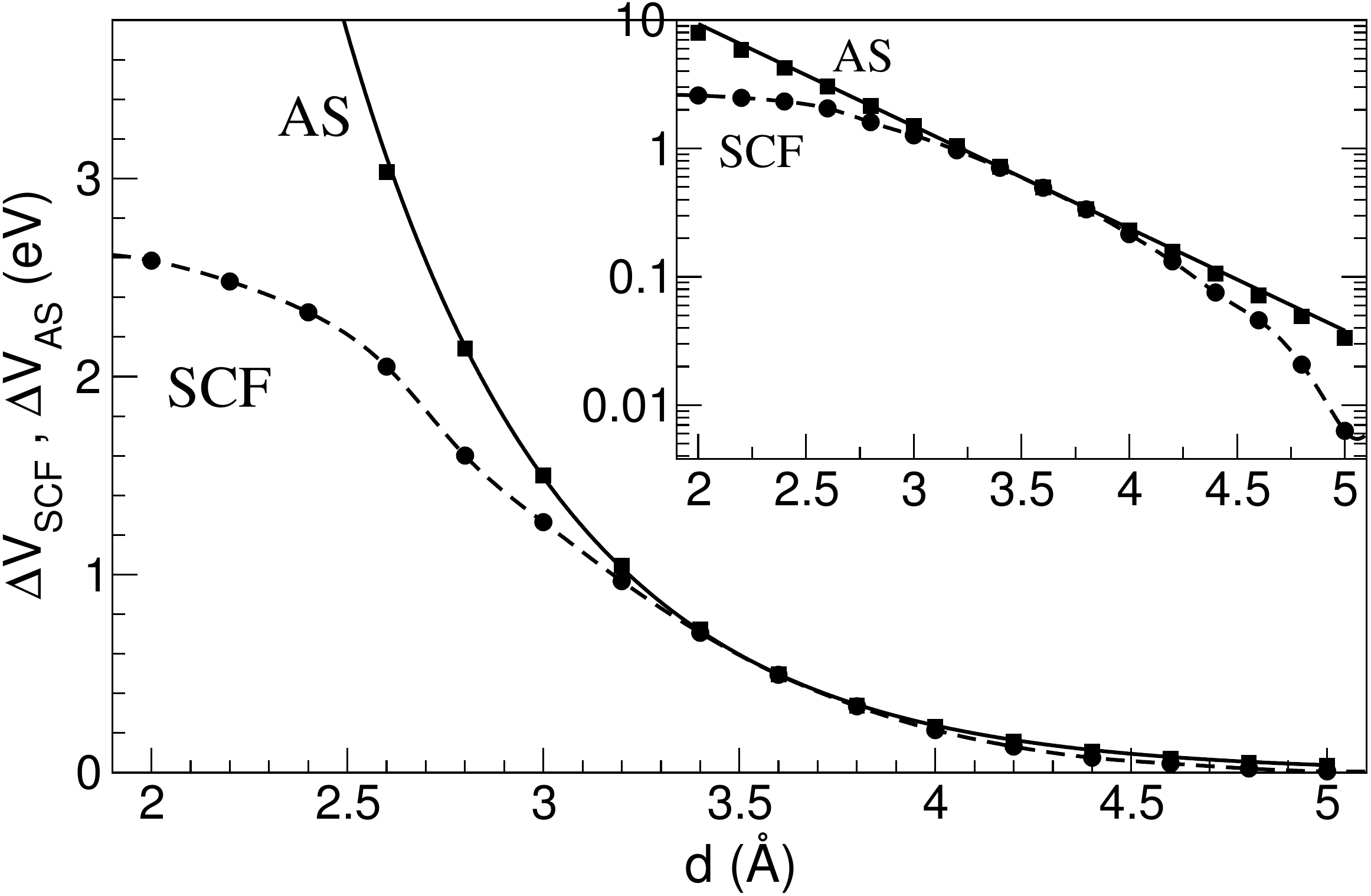}
\caption{
Potential steps $\Delta V_{\rm SCF}$, $\Delta V_{\rm AS}$ at the 
Cu(111)$|$\BN{} 
interface from self-consistent LDA calculations and from the AS 
state, Eq. (\ref{eq:1}), respectively. Inset: potential steps on a logarithmic 
scale.
}
\label{fig3}
\end{figure}

{\color{red}\it Cu(111)$|$\BN{} interface.}---Figure \ref{fig3} shows $\Delta 
V_{\rm AS}$ calculated for the Cu(111)$|$\BN{} interface with the AS state 
constructed as discussed above as a function of $d$, the separation between the 
Cu(111) surface and the \BN{} plane. $\Delta V_{\rm AS}$ is an exponential 
function of $d$, consistent with the fact that it depends on the overlap between 
the Cu and the \BN{} wave functions at the interface. The behavior of the 
potential step obtained from fully self-consistent calculations, $\Delta V_{\rm 
SCF}$, is slightly more complicated.  For separations $d \geq 3.4$ \AA, $\Delta 
V_{\rm SCF}$ coincides with $\Delta V_{\rm AS}$. In this regime exchange 
repulsion between Cu(111) and \BN{} provides a good description of the interface 
potential step.

For separations $d \lesssim 3.4$ \AA , $\Delta V_{\rm SCF}$ deviates from 
$\Delta V_{\rm AS}$. At these shorter distances, stronger (chemical) 
interactions between the two systems become dominant, resulting in a more 
drastic change of the electronic states and in departure from simple exponential 
behavior. At the vdW-DF equilibrium separation, $d_{eq}=3.3$\AA , the potential 
step calculated from exchange repulsion is only $\sim 5$\% higher than the SCF 
value. At the LDA equilibrium separation, $d_{\rm eq}=3.0$\AA , exchange 
repulsion overestimates the SCF value by $\sim 25$\%. 
For distances $d \gtrsim 4.0$ \AA\ $\Delta V_{\rm SCF}$ also starts to deviate 
from $\Delta V_{\rm AS}$. This is caused by the long range vdW interaction 
between  Cu(111) and \BN{}. Note however that at these distances $\Delta V_{{\rm 
SCF}} < 0.1$ eV, so the impact of the long range interactions is in absolute 
terms small.

The plane-averaged electron displacement calculated with the AS 
state, $\Delta\overline{n}_{\rm AS}(z)$, is plotted in Fig.~\ref{fig2} (c)-(d). 
Exchange repulsion pushes electrons out of the overlap region between the 
Cu(111) surface and the \BN{} plane. The system as a whole stays neutral, and 
the depleted electrons are accumulated close to the Cu(111) surface and the 
\BN{} plane. The depletion/accumulation pattern is asymmetric. The wave 
functions of the Cu(111) surface extend more into the vacuum than the \BN{} wave 
functions, implying that the overlap affects the former over a larger region 
than the latter,  and that the effects of  exchange repulsion are larger on the 
Cu(111) side than on the \BN{} side. 

This asymmetric depletion/accumulation pattern results in a net interface dipole 
that points out of the Cu(111) surface. Compared to the clean Cu(111) surface, 
adsorption of \BN{} pushes back some of the electrons that would otherwise spill 
out into the vacuum. The potential step $\Delta V_{\rm AS}$ is downwards going 
from Cu(111) to \BN{}, so exchange repulsion {\em reduces} the work function 
with respect to the clean metal. Such a decrease is commonly found, not only in 
the physisorption of \BN{} on other metal substrates (see below), but also in 
the physisorption of graphene, and of organic molecules. The shape of the 
electron displacement $\Delta\overline{n}_{\rm AS}$ 
depends only weakly on the separation 
$d$ of the \BN{} plane from the Cu(111) surface while its amplitude decreases 
exponentially with increasing separation. 

Exchange repulsion accounts for most of the potential step at distances around 
the equilibrium separation, but $\Delta\overline{n}_{\rm AS}$ is 
not identical to 
$\Delta\overline{n}_{\rm SCF}$, see Fig.~\ref{fig2} (c)-(d). The difference 
$\Delta\overline{n}_{\rm diff} = \Delta\overline{n}_{\rm SCF} - 
\Delta\overline{n}_{\rm AS} =\overline{n}_{\rm SCF} - \overline{n}_{\rm AS}$ measures how the 
orbitals change as a result of chemical and vdW interactions. For all 
separations, $\Delta\overline{n}_{\rm diff}$ describes an accumulation of 
electrons between the Cu(111) surface and the \BN{} plane accompanied by a 
depletion of electrons in the Cu(111) surface and the \BN{} plane, see Fig. 
\ref{fig2}(c)-(d). Such a depletion/accumulation pattern is typical of bond 
formation.  At short distances, $d < 3$ \AA, $\Delta\overline{n}_{\rm diff}$ 
gives a sizable dipole opposite to that calculated from exchange repulsion. 
Interpreting $\Delta\overline{n}_{\rm diff}$ as bond formation, the polarity of 
the bond is then such that \BN{} is on the negative side, which is consistent 
with the fact that \BN{} is more electronegative than Cu. The result is that  
$\Delta V_{{\rm SCF}} < \Delta V_{\rm AS}$.

Remarkably, at distances around the equilibrium separation 3.3 \AA, 
$\Delta\overline{n}_{\rm diff}$ shows a pattern that is fairly symmetric with 
respect to Cu(111) and \BN{}, such that the resulting dipole is moderate and 
results in $\Delta V_{{\rm SCF}} \approx \Delta V$. As $\Delta\overline{n}_{\rm AS}$ goes 
to zero exponentially as a function of $d$, $\Delta\overline{n}_{\rm diff}$ 
approaches $\Delta\overline{n}_{{\rm SCF}}$ for large $d$. The electron 
displacement coming from the vdW bond is the only term remaining at these 
distances, but it yields only a small potential step. 


\begin{figure}[t]
\includegraphics[width=8.5cm]{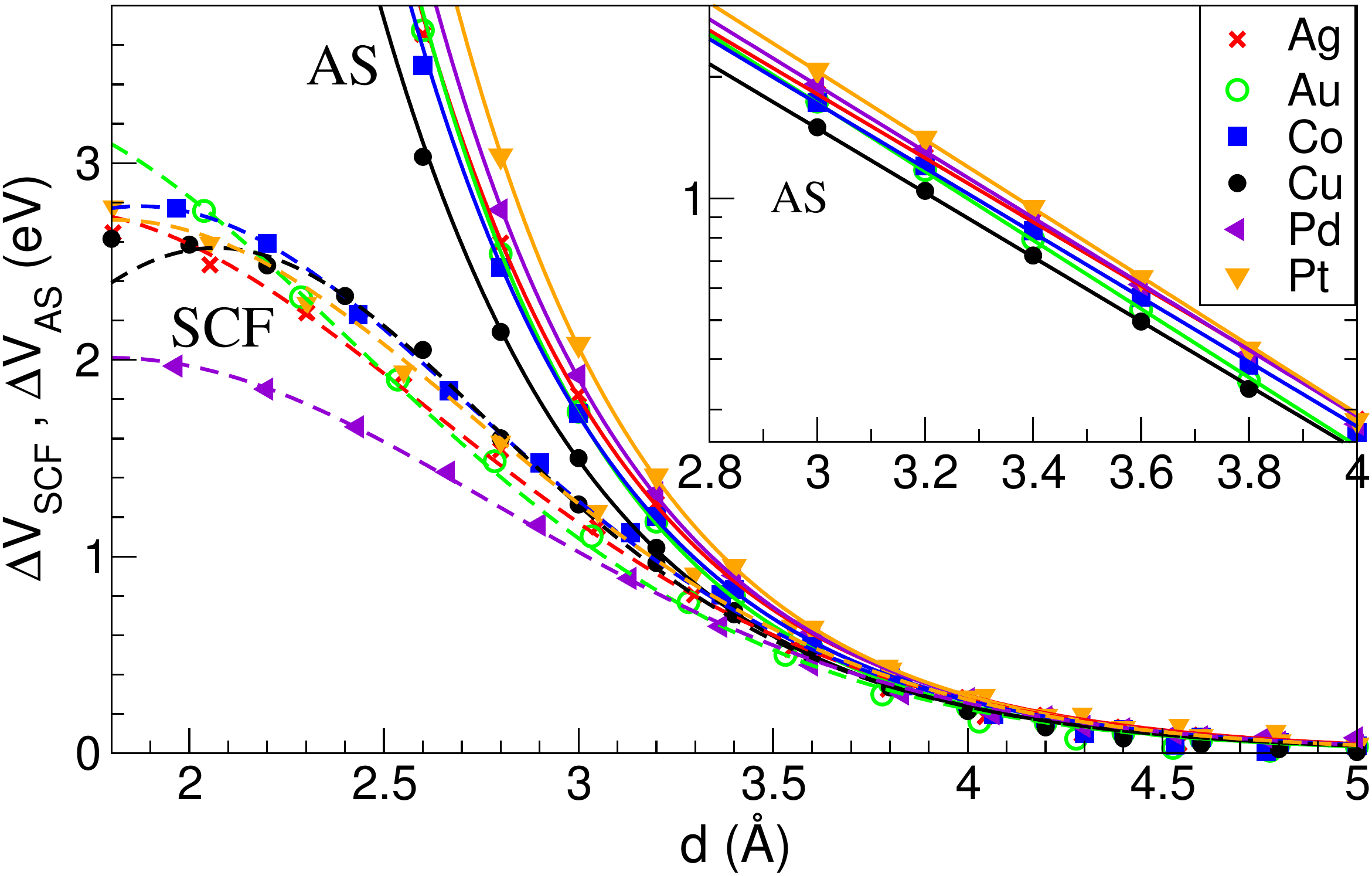}
\caption{(color online) Potential steps $\Delta V_{\rm SCF}$ and $\Delta V_{\rm 
AS}$ at metal$|$\BN{} interfaces from self-consistent LDA calculations and the 
corresponding AS states, respectively. Inset: $\Delta V_{\rm AS}$ on a 
logarithmic scale.}
\label{fig4}
\end{figure}

{\color{red}\it Interface potential steps.}---Fig. \ref{fig4} shows the 
potential steps as a function of $d$ at metal$|$\BN{} interfaces for six 
different metal substrates. At $d \approx 3.5$ \AA\ the curves for the 
self-consistent potential steps, $\Delta V_{\rm SCF}$, converge with those of 
the exchange repulsion potential steps, $\Delta V_{\rm AS}$.  At such 
separations the exchange repulsion is the dominant contribution to the interface 
potential steps. Although the electron displacement coming from the vdW bond has 
the longest range, it does not yield a sizable potential step. At $d<3.0$ \AA\ 
interactions become stronger, and the contribution to the potential step of the 
electron displacement resulting from chemical bonds is not negligible. Compared 
to exchange repulsion only, this contribution tempers the potential steps for 
all metals considered.

The exchange repulsion potential steps exhibit an exponential behavior, $\Delta 
V_{\rm AS}(d)\approx a_0 e^{-\gamma d}$, see the inset to Fig.~\ref{fig4}. 
There is a correlation between the exponent $\gamma$ and the work function $W$ 
of the metal, i.e., $\gamma$ increases if $W$ increases. The correlation is 
weak, however, with $\gamma$ varying between 1.82 for Ag and 1.95 for Pt. It is 
then not surprising that a single, average $\gamma$ gives a reasonable fit for 
the curves of all metals. As the exchange repulsion gives the largest 
contribution to it, we can express the self-consistent potential step about the 
equilibrium separation as $\Delta V_{{\rm SCF}} \approx f(d) e^{-\gamma d}$ 
where $f(d)$ can be described by a simple polynomial 
\cite{Khomyakov:prb09,Bokdam:prb13}. 

{\color{red}\it Summary.}---We have explored the formation of potential steps at 
metal$|$insulator interfaces, using metal$|$\BN{} interfaces as archetypal 
example. Such potential steps can be surprisingly large, i.e., in excess of 1 
eV, even when the bonding is weak, van der Waals bonding. Constructing a model 
for the Pauli exchange repulsion at the interface, we identify the major 
contributions to the interface potential steps. For metal-insulator separations 
that are typical for physisorption, exchange repulsion is the main origin of the 
interface potential step. At these and larger separations, van der Waals 
interactions are important to describe bonding, but give a relatively small 
contribution to the potential step. At shorter distances chemical bonding 
interactions tend to reduce the interface potential step.

{\color{red}\it Acknowledgment.}---M.B. acknowledges support from the 
European project MINOTOR, Grant No. FP7-NMP-228424. The use of supercomputer 
facilities was sponsored by the Physical Sciences (Exacte Wetenschappen) 
division of NWO (Nederlandse Organisatie voor Wetenschappelijk 
Onderzoek).


\begin{thebibliography}{46}%
\makeatletter
\providecommand \@ifxundefined [1]{%
 \@ifx{#1\undefined}
}%
\providecommand \@ifnum [1]{%
 \ifnum #1\expandafter \@firstoftwo
 \else \expandafter \@secondoftwo
 \fi
}%
\providecommand \@ifx [1]{%
 \ifx #1\expandafter \@firstoftwo
 \else \expandafter \@secondoftwo
 \fi
}%
\providecommand \natexlab [1]{#1}%
\providecommand \enquote  [1]{``#1''}%
\providecommand \bibnamefont  [1]{#1}%
\providecommand \bibfnamefont [1]{#1}%
\providecommand \citenamefont [1]{#1}%
\providecommand \href@noop [0]{\@secondoftwo}%
\providecommand \href [0]{\begingroup \@sanitize@url \@href}%
\providecommand \@href[1]{\@@startlink{#1}\@@href}%
\providecommand \@@href[1]{\endgroup#1\@@endlink}%
\providecommand \@sanitize@url [0]{\catcode `\\12\catcode `\$12\catcode
  `\&12\catcode `\#12\catcode `\^12\catcode `\_12\catcode `\%12\relax}%
\providecommand \@@startlink[1]{}%
\providecommand \@@endlink[0]{}%
\providecommand \url  [0]{\begingroup\@sanitize@url \@url }%
\providecommand \@url [1]{\endgroup\@href {#1}{\urlprefix }}%
\providecommand \urlprefix  [0]{URL }%
\providecommand \Eprint [0]{\href }%
\providecommand \doibase [0]{http://dx.doi.org/}%
\providecommand \selectlanguage [0]{\@gobble}%
\providecommand \bibinfo  [0]{\@secondoftwo}%
\providecommand \bibfield  [0]{\@secondoftwo}%
\providecommand \translation [1]{[#1]}%
\providecommand \BibitemOpen [0]{}%
\providecommand \bibitemStop [0]{}%
\providecommand \bibitemNoStop [0]{.\EOS\space}%
\providecommand \EOS [0]{\spacefactor3000\relax}%
\providecommand \BibitemShut  [1]{\csname bibitem#1\endcsname}%
\let\auto@bib@innerbib\@empty
\bibitem [{\citenamefont {Tung}(2014)}]{Tung:apr14}%
  \BibitemOpen
  \bibfield  {author} {\bibinfo {author} {\bibfnamefont {R.~T.}\ \bibnamefont
  {Tung}},\ }\href {\doibase 10.1063/1.4858400} {\bibfield  {journal} {\bibinfo
   {journal} {Appl. Phys. Rev.}\ }\textbf {\bibinfo {volume} {1}},\ \bibinfo
  {pages} {011304} (\bibinfo {year} {2014})}\BibitemShut {NoStop}%
\bibitem [{\citenamefont {Geim}\ and\ \citenamefont
  {Grigorieva}(2013)}]{Geim:nat13}%
  \BibitemOpen
  \bibfield  {author} {\bibinfo {author} {\bibfnamefont {A.~K.}\ \bibnamefont
  {Geim}}\ and\ \bibinfo {author} {\bibfnamefont {I.~V.}\ \bibnamefont
  {Grigorieva}},\ }\href {\doibase 10.1038/nature12385} {\bibfield  {journal}
  {\bibinfo  {journal} {Nature}\ }\textbf {\bibinfo {volume} {499}},\ \bibinfo
  {pages} {419} (\bibinfo {year} {2013})}\BibitemShut {NoStop}%
\bibitem [{\citenamefont {Popov}\ \emph {et~al.}(2012)\citenamefont {Popov},
  \citenamefont {Seifert},\ and\ \citenamefont {Tom{\'{a}}nek}}]{Popov:prl12}%
  \BibitemOpen
  \bibfield  {author} {\bibinfo {author} {\bibfnamefont {I.}~\bibnamefont
  {Popov}}, \bibinfo {author} {\bibfnamefont {G.}~\bibnamefont {Seifert}}, \
  and\ \bibinfo {author} {\bibfnamefont {D.}~\bibnamefont {Tom{\'{a}}nek}},\
  }\href {\doibase 10.1103/PhysRevLett.108.156802} {\bibfield  {journal}
  {\bibinfo  {journal} {Phys. Rev. Lett.}\ }\textbf {\bibinfo {volume} {108}},\
  \bibinfo {pages} {156802} (\bibinfo {year} {2012})}\BibitemShut {NoStop}%
\bibitem [{\citenamefont {Chen}\ \emph {et~al.}(2013)\citenamefont {Chen},
  \citenamefont {Santos}, \citenamefont {Zhu}, \citenamefont {Kaxiras},\ and\
  \citenamefont {Zhang}}]{Chen:nanol13}%
  \BibitemOpen
  \bibfield  {author} {\bibinfo {author} {\bibfnamefont {W.}~\bibnamefont
  {Chen}}, \bibinfo {author} {\bibfnamefont {E.~J.~G.}\ \bibnamefont {Santos}},
  \bibinfo {author} {\bibfnamefont {W.}~\bibnamefont {Zhu}}, \bibinfo {author}
  {\bibfnamefont {E.}~\bibnamefont {Kaxiras}}, \ and\ \bibinfo {author}
  {\bibfnamefont {Z.}~\bibnamefont {Zhang}},\ }\href {\doibase
  10.1021/nl303909f} {\bibfield  {journal} {\bibinfo  {journal} {Nano Letters}\
  }\textbf {\bibinfo {volume} {13}},\ \bibinfo {pages} {509} (\bibinfo {year}
  {2013})}\BibitemShut {NoStop}%
\bibitem [{\citenamefont {Gong}\ \emph {et~al.}(2014)\citenamefont {Gong},
  \citenamefont {Colombo}, \citenamefont {Wallace},\ and\ \citenamefont
  {Cho}}]{Gong:nanol14}%
  \BibitemOpen
  \bibfield  {author} {\bibinfo {author} {\bibfnamefont {C.}~\bibnamefont
  {Gong}}, \bibinfo {author} {\bibfnamefont {L.}~\bibnamefont {Colombo}},
  \bibinfo {author} {\bibfnamefont {R.~M.}\ \bibnamefont {Wallace}}, \ and\
  \bibinfo {author} {\bibfnamefont {K.}~\bibnamefont {Cho}},\ }\href {\doibase
  10.1021/nl403465v} {\bibfield  {journal} {\bibinfo  {journal} {Nano Letters}\
  }\textbf {\bibinfo {volume} {14}},\ \bibinfo {pages} {1714} (\bibinfo {year}
  {2014})}\BibitemShut {NoStop}%
\bibitem [{\citenamefont {Kang}\ \emph {et~al.}(2014)\citenamefont {Kang},
  \citenamefont {Liu}, \citenamefont {Sarkar}, \citenamefont {Jena},\ and\
  \citenamefont {Banerjee}}]{Kang:prx14}%
  \BibitemOpen
  \bibfield  {author} {\bibinfo {author} {\bibfnamefont {J.}~\bibnamefont
  {Kang}}, \bibinfo {author} {\bibfnamefont {W.}~\bibnamefont {Liu}}, \bibinfo
  {author} {\bibfnamefont {D.}~\bibnamefont {Sarkar}}, \bibinfo {author}
  {\bibfnamefont {D.}~\bibnamefont {Jena}}, \ and\ \bibinfo {author}
  {\bibfnamefont {K.}~\bibnamefont {Banerjee}},\ }\href {\doibase
  10.1103/PhysRevX.4.031005} {\bibfield  {journal} {\bibinfo  {journal} {Phys.
  Rev. X}\ }\textbf {\bibinfo {volume} {4}},\ \bibinfo {pages} {031005}
  (\bibinfo {year} {2014})}\BibitemShut {NoStop}%
\bibitem [{\citenamefont {Dean}\ \emph {et~al.}(2010)\citenamefont {Dean},
  \citenamefont {Young}, \citenamefont {Meric}, \citenamefont {Lee},
  \citenamefont {Wang}, \citenamefont {Sorgenfrei}, \citenamefont {Watanabe},
  \citenamefont {Taniguchi}, \citenamefont {Kim}, \citenamefont {Shepard},\
  and\ \citenamefont {Hone}}]{Dean:natn10}%
  \BibitemOpen
  \bibfield  {author} {\bibinfo {author} {\bibfnamefont {C.~R.}\ \bibnamefont
  {Dean}}, \bibinfo {author} {\bibfnamefont {A.~F.}\ \bibnamefont {Young}},
  \bibinfo {author} {\bibfnamefont {I.}~\bibnamefont {Meric}}, \bibinfo
  {author} {\bibfnamefont {C.}~\bibnamefont {Lee}}, \bibinfo {author}
  {\bibfnamefont {L.}~\bibnamefont {Wang}}, \bibinfo {author} {\bibfnamefont
  {S.}~\bibnamefont {Sorgenfrei}}, \bibinfo {author} {\bibfnamefont
  {K.}~\bibnamefont {Watanabe}}, \bibinfo {author} {\bibfnamefont
  {T.}~\bibnamefont {Taniguchi}}, \bibinfo {author} {\bibfnamefont
  {P.}~\bibnamefont {Kim}}, \bibinfo {author} {\bibfnamefont {K.~L.}\
  \bibnamefont {Shepard}}, \ and\ \bibinfo {author} {\bibfnamefont
  {J.}~\bibnamefont {Hone}},\ }\href {\doibase 10.1038/nnano.2010.172}
  {\bibfield  {journal} {\bibinfo  {journal} {Nature Nanotechnology}\ }\textbf
  {\bibinfo {volume} {5}},\ \bibinfo {pages} {722} (\bibinfo {year}
  {2010})}\BibitemShut {NoStop}%
\bibitem [{\citenamefont {Nagashima}\ \emph {et~al.}(1996)\citenamefont
  {Nagashima}, \citenamefont {Tejima}, \citenamefont {Gamou}, \citenamefont
  {Kawai},\ and\ \citenamefont {Oshima}}]{Nagashima:ss96}%
  \BibitemOpen
  \bibfield  {author} {\bibinfo {author} {\bibfnamefont {A.}~\bibnamefont
  {Nagashima}}, \bibinfo {author} {\bibfnamefont {N.}~\bibnamefont {Tejima}},
  \bibinfo {author} {\bibfnamefont {Y.}~\bibnamefont {Gamou}}, \bibinfo
  {author} {\bibfnamefont {T.}~\bibnamefont {Kawai}}, \ and\ \bibinfo {author}
  {\bibfnamefont {C.}~\bibnamefont {Oshima}},\ }\href {\doibase
  10.1016/0039-6028(96)00134-3} {\bibfield  {journal} {\bibinfo  {journal}
  {Surface Science}\ }\textbf {\bibinfo {volume} {357-358}},\ \bibinfo {pages}
  {307} (\bibinfo {year} {1996})}\BibitemShut {NoStop}%
\bibitem [{\citenamefont {Preobrajenski}\ \emph {et~al.}(2005)\citenamefont
  {Preobrajenski}, \citenamefont {Vinogradov},\ and\ \citenamefont
  {M{\aa}rtensson}}]{Preobrajenski:ss05}%
  \BibitemOpen
  \bibfield  {author} {\bibinfo {author} {\bibfnamefont {A.~B.}\ \bibnamefont
  {Preobrajenski}}, \bibinfo {author} {\bibfnamefont {A.~S.}\ \bibnamefont
  {Vinogradov}}, \ and\ \bibinfo {author} {\bibfnamefont {N.}~\bibnamefont
  {M{\aa}rtensson}},\ }\href {\doibase 10.1016/j.susc.2005.02.047} {\bibfield
  {journal} {\bibinfo  {journal} {Surface Science}\ }\textbf {\bibinfo {volume}
  {582}},\ \bibinfo {pages} {21} (\bibinfo {year} {2005})}\BibitemShut
  {NoStop}%
\bibitem [{\citenamefont {Leuenberger}\ \emph {et~al.}(2011)\citenamefont
  {Leuenberger}, \citenamefont {Yanagisawa}, \citenamefont {Roth},
  \citenamefont {Osterwalder},\ and\ \citenamefont
  {Hengsberger}}]{Leuenberger:prb11}%
  \BibitemOpen
  \bibfield  {author} {\bibinfo {author} {\bibfnamefont {D.}~\bibnamefont
  {Leuenberger}}, \bibinfo {author} {\bibfnamefont {H.}~\bibnamefont
  {Yanagisawa}}, \bibinfo {author} {\bibfnamefont {S.}~\bibnamefont {Roth}},
  \bibinfo {author} {\bibfnamefont {J.}~\bibnamefont {Osterwalder}}, \ and\
  \bibinfo {author} {\bibfnamefont {M.}~\bibnamefont {Hengsberger}},\ }\href
  {\doibase 10.1103/PhysRevB.84.125107} {\bibfield  {journal} {\bibinfo
  {journal} {Phys. Rev. B}\ }\textbf {\bibinfo {volume} {84}},\ \bibinfo
  {pages} {125107} (\bibinfo {year} {2011})}\BibitemShut {NoStop}%
\bibitem [{\citenamefont {Joshi}\ \emph {et~al.}(2012)\citenamefont {Joshi},
  \citenamefont {Ecija}, \citenamefont {Koitz}, \citenamefont {Iannuzzi},
  \citenamefont {Seitsonen}, \citenamefont {Hutter}, \citenamefont {Sachdev},
  \citenamefont {Vijayaraghavan}, \citenamefont {Bischoff}, \citenamefont
  {Seufert}, \citenamefont {Barth},\ and\ \citenamefont
  {Auw{\"{a}}rter}}]{Joshi:nanol12}%
  \BibitemOpen
  \bibfield  {author} {\bibinfo {author} {\bibfnamefont {S.}~\bibnamefont
  {Joshi}}, \bibinfo {author} {\bibfnamefont {D.}~\bibnamefont {Ecija}},
  \bibinfo {author} {\bibfnamefont {R.}~\bibnamefont {Koitz}}, \bibinfo
  {author} {\bibfnamefont {M.}~\bibnamefont {Iannuzzi}}, \bibinfo {author}
  {\bibfnamefont {A.~P.}\ \bibnamefont {Seitsonen}}, \bibinfo {author}
  {\bibfnamefont {J.}~\bibnamefont {Hutter}}, \bibinfo {author} {\bibfnamefont
  {H.}~\bibnamefont {Sachdev}}, \bibinfo {author} {\bibfnamefont
  {S.}~\bibnamefont {Vijayaraghavan}}, \bibinfo {author} {\bibfnamefont
  {F.}~\bibnamefont {Bischoff}}, \bibinfo {author} {\bibfnamefont
  {K.}~\bibnamefont {Seufert}}, \bibinfo {author} {\bibfnamefont {J.~V.}\
  \bibnamefont {Barth}}, \ and\ \bibinfo {author} {\bibfnamefont
  {W.}~\bibnamefont {Auw{\"{a}}rter}},\ }\href {\doibase 10.1021/nl303170m}
  {\bibfield  {journal} {\bibinfo  {journal} {Nano Letters}\ }\textbf {\bibinfo
  {volume} {12}},\ \bibinfo {pages} {5821} (\bibinfo {year}
  {2012})}\BibitemShut {NoStop}%
\bibitem [{\citenamefont {Olsen}\ \emph {et~al.}(2011)\citenamefont {Olsen},
  \citenamefont {Yan}, \citenamefont {Mortensen},\ and\ \citenamefont
  {Thygesen}}]{Olsen:prl11}%
  \BibitemOpen
  \bibfield  {author} {\bibinfo {author} {\bibfnamefont {T.}~\bibnamefont
  {Olsen}}, \bibinfo {author} {\bibfnamefont {J.}~\bibnamefont {Yan}}, \bibinfo
  {author} {\bibfnamefont {J.~J.}\ \bibnamefont {Mortensen}}, \ and\ \bibinfo
  {author} {\bibfnamefont {K.~S.}\ \bibnamefont {Thygesen}},\ }\href {\doibase
  10.1103/PhysRevLett.107.156401} {\bibfield  {journal} {\bibinfo  {journal}
  {Phys. Rev. Lett.}\ }\textbf {\bibinfo {volume} {107}},\ \bibinfo {pages}
  {156401} (\bibinfo {year} {2011})}\BibitemShut {NoStop}%
\bibitem [{\citenamefont {Stradi}\ \emph {et~al.}(2011)\citenamefont {Stradi},
  \citenamefont {Barja}, \citenamefont {D{\'{i}}az}, \citenamefont {Garnica},
  \citenamefont {Borca}, \citenamefont {Hinarejos}, \citenamefont
  {S{\'{a}}nchez-Portal}, \citenamefont {Alcam{\'{i}}}, \citenamefont {Arnau},
  \citenamefont {{V{\'{a}}zquez de Parga}}, \citenamefont {Miranda},\ and\
  \citenamefont {Mart{\'{i}}n}}]{Stradi:prl11}%
  \BibitemOpen
  \bibfield  {author} {\bibinfo {author} {\bibfnamefont {D.}~\bibnamefont
  {Stradi}}, \bibinfo {author} {\bibfnamefont {S.}~\bibnamefont {Barja}},
  \bibinfo {author} {\bibfnamefont {C.}~\bibnamefont {D{\'{i}}az}}, \bibinfo
  {author} {\bibfnamefont {M.}~\bibnamefont {Garnica}}, \bibinfo {author}
  {\bibfnamefont {B.}~\bibnamefont {Borca}}, \bibinfo {author} {\bibfnamefont
  {J.~J.}\ \bibnamefont {Hinarejos}}, \bibinfo {author} {\bibfnamefont
  {D.}~\bibnamefont {S{\'{a}}nchez-Portal}}, \bibinfo {author} {\bibfnamefont
  {M.}~\bibnamefont {Alcam{\'{i}}}}, \bibinfo {author} {\bibfnamefont
  {A.}~\bibnamefont {Arnau}}, \bibinfo {author} {\bibfnamefont {A.~L.}\
  \bibnamefont {{V{\'{a}}zquez de Parga}}}, \bibinfo {author} {\bibfnamefont
  {R.}~\bibnamefont {Miranda}}, \ and\ \bibinfo {author} {\bibfnamefont
  {F.}~\bibnamefont {Mart{\'{i}}n}},\ }\href {\doibase
  10.1103/PhysRevLett.106.186102} {\bibfield  {journal} {\bibinfo  {journal}
  {Phys. Rev. Lett.}\ }\textbf {\bibinfo {volume} {106}},\ \bibinfo {pages}
  {186102} (\bibinfo {year} {2011})}\BibitemShut {NoStop}%
\bibitem [{\citenamefont {Giovannetti}\ \emph {et~al.}(2008)\citenamefont
  {Giovannetti}, \citenamefont {Khomyakov}, \citenamefont {Brocks},
  \citenamefont {Karpan}, \citenamefont {van~den Brink},\ and\ \citenamefont
  {Kelly}}]{Giovannetti:prl08}%
  \BibitemOpen
  \bibfield  {author} {\bibinfo {author} {\bibfnamefont {G.}~\bibnamefont
  {Giovannetti}}, \bibinfo {author} {\bibfnamefont {P.~A.}\ \bibnamefont
  {Khomyakov}}, \bibinfo {author} {\bibfnamefont {G.}~\bibnamefont {Brocks}},
  \bibinfo {author} {\bibfnamefont {V.~M.}\ \bibnamefont {Karpan}}, \bibinfo
  {author} {\bibfnamefont {J.}~\bibnamefont {van~den Brink}}, \ and\ \bibinfo
  {author} {\bibfnamefont {P.~J.}\ \bibnamefont {Kelly}},\ }\href {\doibase
  10.1103/PhysRevLett.101.026803} {\bibfield  {journal} {\bibinfo  {journal}
  {Phys. Rev. Lett.}\ }\textbf {\bibinfo {volume} {101}},\ \bibinfo {pages}
  {026803} (\bibinfo {year} {2008})}\BibitemShut {NoStop}%
\bibitem [{\citenamefont {Khomyakov}\ \emph {et~al.}(2009)\citenamefont
  {Khomyakov}, \citenamefont {Giovannetti}, \citenamefont {Rusu}, \citenamefont
  {Brocks}, \citenamefont {van~den Brink},\ and\ \citenamefont
  {Kelly}}]{Khomyakov:prb09}%
  \BibitemOpen
  \bibfield  {author} {\bibinfo {author} {\bibfnamefont {P.~A.}\ \bibnamefont
  {Khomyakov}}, \bibinfo {author} {\bibfnamefont {G.}~\bibnamefont
  {Giovannetti}}, \bibinfo {author} {\bibfnamefont {P.~C.}\ \bibnamefont
  {Rusu}}, \bibinfo {author} {\bibfnamefont {G.}~\bibnamefont {Brocks}},
  \bibinfo {author} {\bibfnamefont {J.}~\bibnamefont {van~den Brink}}, \ and\
  \bibinfo {author} {\bibfnamefont {P.~J.}\ \bibnamefont {Kelly}},\ }\href
  {\doibase 10.1103/PhysRevB.79.195425} {\bibfield  {journal} {\bibinfo
  {journal} {Phys. Rev. B}\ }\textbf {\bibinfo {volume} {79}},\ \bibinfo
  {pages} {195425} (\bibinfo {year} {2009})}\BibitemShut {NoStop}%
\bibitem [{\citenamefont {Bokdam}\ \emph {et~al.}(2011)\citenamefont {Bokdam},
  \citenamefont {Khomyakov}, \citenamefont {Brocks}, \citenamefont {Zhong},\
  and\ \citenamefont {Kelly}}]{Bokdam:nanol11}%
  \BibitemOpen
  \bibfield  {author} {\bibinfo {author} {\bibfnamefont {M.}~\bibnamefont
  {Bokdam}}, \bibinfo {author} {\bibfnamefont {P.~A.}\ \bibnamefont
  {Khomyakov}}, \bibinfo {author} {\bibfnamefont {G.}~\bibnamefont {Brocks}},
  \bibinfo {author} {\bibfnamefont {Z.}~\bibnamefont {Zhong}}, \ and\ \bibinfo
  {author} {\bibfnamefont {P.~J.}\ \bibnamefont {Kelly}},\ }\href {\doibase
  10.1021/nl202131q} {\bibfield  {journal} {\bibinfo  {journal} {Nano Letters}\
  }\textbf {\bibinfo {volume} {11}},\ \bibinfo {pages} {4631} (\bibinfo {year}
  {2011})}\BibitemShut {NoStop}%
\bibitem [{\citenamefont {Bokdam}\ \emph {et~al.}(2013)\citenamefont {Bokdam},
  \citenamefont {Khomyakov}, \citenamefont {Brocks},\ and\ \citenamefont
  {Kelly}}]{Bokdam:prb13}%
  \BibitemOpen
  \bibfield  {author} {\bibinfo {author} {\bibfnamefont {M.}~\bibnamefont
  {Bokdam}}, \bibinfo {author} {\bibfnamefont {P.~A.}\ \bibnamefont
  {Khomyakov}}, \bibinfo {author} {\bibfnamefont {G.}~\bibnamefont {Brocks}}, \
  and\ \bibinfo {author} {\bibfnamefont {P.~J.}\ \bibnamefont {Kelly}},\ }\href
  {\doibase 10.1103/PhysRevB.87.075414} {\bibfield  {journal} {\bibinfo
  {journal} {Phys. Rev. B}\ }\textbf {\bibinfo {volume} {87}},\ \bibinfo
  {pages} {075414} (\bibinfo {year} {2013})}\BibitemShut {NoStop}%
\bibitem [{\citenamefont {Gebhardt}\ \emph {et~al.}(2012)\citenamefont
  {Gebhardt}, \citenamefont {Vi{\~{n}}es},\ and\ \citenamefont
  {G{\"{o}}rling}}]{Gebhardt:prb12}%
  \BibitemOpen
  \bibfield  {author} {\bibinfo {author} {\bibfnamefont {J.}~\bibnamefont
  {Gebhardt}}, \bibinfo {author} {\bibfnamefont {F.}~\bibnamefont
  {Vi{\~{n}}es}}, \ and\ \bibinfo {author} {\bibfnamefont {A.}~\bibnamefont
  {G{\"{o}}rling}},\ }\href {\doibase 10.1103/PhysRevB.86.195431} {\bibfield
  {journal} {\bibinfo  {journal} {Phys. Rev. B}\ }\textbf {\bibinfo {volume}
  {86}},\ \bibinfo {pages} {195431} (\bibinfo {year} {2012})}\BibitemShut
  {NoStop}%
\bibitem [{\citenamefont {Nagashima}\ \emph {et~al.}(1995)\citenamefont
  {Nagashima}, \citenamefont {Tejima}, \citenamefont {Gamou}, \citenamefont
  {Kawai},\ and\ \citenamefont {Oshima}}]{Nagashima:prl95}%
  \BibitemOpen
  \bibfield  {author} {\bibinfo {author} {\bibfnamefont {A.}~\bibnamefont
  {Nagashima}}, \bibinfo {author} {\bibfnamefont {N.}~\bibnamefont {Tejima}},
  \bibinfo {author} {\bibfnamefont {Y.}~\bibnamefont {Gamou}}, \bibinfo
  {author} {\bibfnamefont {T.}~\bibnamefont {Kawai}}, \ and\ \bibinfo {author}
  {\bibfnamefont {C.}~\bibnamefont {Oshima}},\ }\href {\doibase
  10.1103/PhysRevLett.75.3918} {\bibfield  {journal} {\bibinfo  {journal}
  {Phys. Rev. Lett.}\ }\textbf {\bibinfo {volume} {75}},\ \bibinfo {pages}
  {3918} (\bibinfo {year} {1995})}\BibitemShut {NoStop}%
\bibitem [{\citenamefont {Schulz}\ \emph {et~al.}(2014)\citenamefont {Schulz},
  \citenamefont {Drost}, \citenamefont {H{\"{a}}m{\"{a}}l{\"{a}}inen},
  \citenamefont {Demonchaux}, \citenamefont {Seitsonen},\ and\ \citenamefont
  {Liljeroth}}]{Schulz:prb14}%
  \BibitemOpen
  \bibfield  {author} {\bibinfo {author} {\bibfnamefont {F.}~\bibnamefont
  {Schulz}}, \bibinfo {author} {\bibfnamefont {R.}~\bibnamefont {Drost}},
  \bibinfo {author} {\bibfnamefont {S.~K.}\ \bibnamefont
  {H{\"{a}}m{\"{a}}l{\"{a}}inen}}, \bibinfo {author} {\bibfnamefont
  {T.}~\bibnamefont {Demonchaux}}, \bibinfo {author} {\bibfnamefont {A.~P.}\
  \bibnamefont {Seitsonen}}, \ and\ \bibinfo {author} {\bibfnamefont
  {P.}~\bibnamefont {Liljeroth}},\ }\href {\doibase 10.1103/PhysRevB.89.235429}
  {\bibfield  {journal} {\bibinfo  {journal} {Phys. Rev. B}\ }\textbf {\bibinfo
  {volume} {89}},\ \bibinfo {pages} {235429} (\bibinfo {year}
  {2014})}\BibitemShut {NoStop}%
\bibitem [{\citenamefont {Brocks}\ \emph {et~al.}(1984)\citenamefont {Brocks},
  \citenamefont {Tennyson},\ and\ \citenamefont {{van der
  Avoird}}}]{Brocks:jcp84}%
  \BibitemOpen
  \bibfield  {author} {\bibinfo {author} {\bibfnamefont {G.}~\bibnamefont
  {Brocks}}, \bibinfo {author} {\bibfnamefont {J.}~\bibnamefont {Tennyson}}, \
  and\ \bibinfo {author} {\bibfnamefont {A.}~\bibnamefont {{van der Avoird}}},\
  }\href {\doibase 10.1063/1.447075} {\bibfield  {journal} {\bibinfo  {journal}
  {J. Chem. Phys.}\ }\textbf {\bibinfo {volume} {80}},\ \bibinfo {pages} {3223}
  (\bibinfo {year} {1984})}\BibitemShut {NoStop}%
\bibitem [{\citenamefont {Brocks}\ and\ \citenamefont {{van der
  Avoird}}(1985)}]{Brocks:molphys85}%
  \BibitemOpen
  \bibfield  {author} {\bibinfo {author} {\bibfnamefont {G.}~\bibnamefont
  {Brocks}}\ and\ \bibinfo {author} {\bibfnamefont {A.}~\bibnamefont {{van der
  Avoird}}},\ }\href {\doibase 10.1080/00268978500101131} {\bibfield  {journal}
  {\bibinfo  {journal} {Mol. Phys.}\ }\textbf {\bibinfo {volume} {55}},\
  \bibinfo {pages} {11} (\bibinfo {year} {1985})}\BibitemShut {NoStop}%
\bibitem [{\citenamefont {Dresser}\ \emph {et~al.}(1974)\citenamefont
  {Dresser}, \citenamefont {Madey},\ and\ \citenamefont
  {Yates}}]{Dresser:ss74}%
  \BibitemOpen
  \bibfield  {author} {\bibinfo {author} {\bibfnamefont {M.~J.}\ \bibnamefont
  {Dresser}}, \bibinfo {author} {\bibfnamefont {T.~E.}\ \bibnamefont {Madey}},
  \ and\ \bibinfo {author} {\bibfnamefont {J.~T.}\ \bibnamefont {Yates}},\
  }\href {\doibase 10.1016/0039-6028(74)90037-5} {\bibfield  {journal}
  {\bibinfo  {journal} {Surface Science}\ }\textbf {\bibinfo {volume} {42}},\
  \bibinfo {pages} {533} (\bibinfo {year} {1974})}\BibitemShut {NoStop}%
\bibitem [{\citenamefont {Bagus}\ \emph {et~al.}(2002)\citenamefont {Bagus},
  \citenamefont {Staemmler},\ and\ \citenamefont {W{\"{o}}ll}}]{Bagus:prl02}%
  \BibitemOpen
  \bibfield  {author} {\bibinfo {author} {\bibfnamefont {P.~S.}\ \bibnamefont
  {Bagus}}, \bibinfo {author} {\bibfnamefont {V.}~\bibnamefont {Staemmler}}, \
  and\ \bibinfo {author} {\bibfnamefont {C.}~\bibnamefont {W{\"{o}}ll}},\
  }\href {\doibase 10.1103/PhysRevLett.89.096104} {\bibfield  {journal}
  {\bibinfo  {journal} {Phys. Rev. Lett.}\ }\textbf {\bibinfo {volume} {89}},\
  \bibinfo {pages} {096104} (\bibinfo {year} {2002})}\BibitemShut {NoStop}%
\bibitem [{\citenamefont {V{\'{a}}zquez}\ \emph {et~al.}(2007)\citenamefont
  {V{\'{a}}zquez}, \citenamefont {Dappe}, \citenamefont {Ortega},\ and\
  \citenamefont {Flores}}]{Vazquez:jcp07}%
  \BibitemOpen
  \bibfield  {author} {\bibinfo {author} {\bibfnamefont {H.}~\bibnamefont
  {V{\'{a}}zquez}}, \bibinfo {author} {\bibfnamefont {Y.~J.}\ \bibnamefont
  {Dappe}}, \bibinfo {author} {\bibfnamefont {J.}~\bibnamefont {Ortega}}, \
  and\ \bibinfo {author} {\bibfnamefont {F.}~\bibnamefont {Flores}},\ }\href
  {\doibase 10.1063/1.2717165} {\bibfield  {journal} {\bibinfo  {journal} {J.
  Chem. Phys.}\ }\textbf {\bibinfo {volume} {126}},\ \bibinfo {pages} {144703}
  (\bibinfo {year} {2007})}\BibitemShut {NoStop}%
\bibitem [{\citenamefont {Bagus}\ \emph {et~al.}(2008)\citenamefont {Bagus},
  \citenamefont {K{\"{a}}fer}, \citenamefont {Witte},\ and\ \citenamefont
  {W{\"{o}}ll}}]{Bagus:prl08}%
  \BibitemOpen
  \bibfield  {author} {\bibinfo {author} {\bibfnamefont {P.~S.}\ \bibnamefont
  {Bagus}}, \bibinfo {author} {\bibfnamefont {D.}~\bibnamefont {K{\"{a}}fer}},
  \bibinfo {author} {\bibfnamefont {G.}~\bibnamefont {Witte}}, \ and\ \bibinfo
  {author} {\bibfnamefont {C.}~\bibnamefont {W{\"{o}}ll}},\ }\href {\doibase
  10.1103/PhysRevLett.100.126101} {\bibfield  {journal} {\bibinfo  {journal}
  {Phys. Rev. Lett.}\ }\textbf {\bibinfo {volume} {100}},\ \bibinfo {pages}
  {126101} (\bibinfo {year} {2008})}\BibitemShut {NoStop}%
\bibitem [{\citenamefont {Vitali}\ \emph {et~al.}(2010)\citenamefont {Vitali},
  \citenamefont {Levita}, \citenamefont {Ohmann}, \citenamefont {Comisso},
  \citenamefont {Vita},\ and\ \citenamefont {Kern}}]{Vitali:natm10}%
  \BibitemOpen
  \bibfield  {author} {\bibinfo {author} {\bibfnamefont {L.}~\bibnamefont
  {Vitali}}, \bibinfo {author} {\bibfnamefont {G.}~\bibnamefont {Levita}},
  \bibinfo {author} {\bibfnamefont {R.}~\bibnamefont {Ohmann}}, \bibinfo
  {author} {\bibfnamefont {A.}~\bibnamefont {Comisso}}, \bibinfo {author}
  {\bibfnamefont {A.~D.}\ \bibnamefont {Vita}}, \ and\ \bibinfo {author}
  {\bibfnamefont {K.}~\bibnamefont {Kern}},\ }\href {\doibase 10.1038/NMAT2625}
  {\bibfield  {journal} {\bibinfo  {journal} {Nature Materials}\ }\textbf
  {\bibinfo {volume} {9}},\ \bibinfo {pages} {320} (\bibinfo {year}
  {2010})}\BibitemShut {NoStop}%
\bibitem [{\citenamefont {Rusu}\ \emph {et~al.}(2010)\citenamefont {Rusu},
  \citenamefont {Giovannetti}, \citenamefont {Weijtens}, \citenamefont
  {Coehoorn},\ and\ \citenamefont {Brocks}}]{Rusu:prb10}%
  \BibitemOpen
  \bibfield  {author} {\bibinfo {author} {\bibfnamefont {P.~C.}\ \bibnamefont
  {Rusu}}, \bibinfo {author} {\bibfnamefont {G.}~\bibnamefont {Giovannetti}},
  \bibinfo {author} {\bibfnamefont {C.}~\bibnamefont {Weijtens}}, \bibinfo
  {author} {\bibfnamefont {R.}~\bibnamefont {Coehoorn}}, \ and\ \bibinfo
  {author} {\bibfnamefont {G.}~\bibnamefont {Brocks}},\ }\href {\doibase
  10.1103/PhysRevB.81.125403} {\bibfield  {journal} {\bibinfo  {journal} {Phys.
  Rev. B}\ }\textbf {\bibinfo {volume} {81}},\ \bibinfo {pages} {125403}
  (\bibinfo {year} {2010})}\BibitemShut {NoStop}%
\bibitem [{\citenamefont {Neugebauer}\ and\ \citenamefont
  {Scheffler}(1992)}]{Neugebauer:prb92}%
  \BibitemOpen
  \bibfield  {author} {\bibinfo {author} {\bibfnamefont {J.}~\bibnamefont
  {Neugebauer}}\ and\ \bibinfo {author} {\bibfnamefont {M.}~\bibnamefont
  {Scheffler}},\ }\href@noop {} {\bibfield  {journal} {\bibinfo  {journal}
  {Phys. Rev. B}\ }\textbf {\bibinfo {volume} {46}},\ \bibinfo {pages} {16067}
  (\bibinfo {year} {1992})}\BibitemShut {NoStop}%
\bibitem [{\citenamefont {{G{\'{o}}mez D{\'{i}}az}}\ \emph
  {et~al.}(2013)\citenamefont {{G{\'{o}}mez D{\'{i}}az}}, \citenamefont {Ding},
  \citenamefont {Koitz}, \citenamefont {Seitsonen}, \citenamefont {Iannuzzi},\
  and\ \citenamefont {Hutter}}]{GomezDiaz:tca13}%
  \BibitemOpen
  \bibfield  {author} {\bibinfo {author} {\bibfnamefont {J.}~\bibnamefont
  {{G{\'{o}}mez D{\'{i}}az}}}, \bibinfo {author} {\bibfnamefont
  {Y.}~\bibnamefont {Ding}}, \bibinfo {author} {\bibfnamefont {R.}~\bibnamefont
  {Koitz}}, \bibinfo {author} {\bibfnamefont {A.~P.}\ \bibnamefont
  {Seitsonen}}, \bibinfo {author} {\bibfnamefont {M.}~\bibnamefont {Iannuzzi}},
  \ and\ \bibinfo {author} {\bibfnamefont {J.}~\bibnamefont {Hutter}},\ }\href
  {\doibase 10.1007/s00214-013-1350-z} {\bibfield  {journal} {\bibinfo
  {journal} {Theor. Chem. Acc.}\ }\textbf {\bibinfo {volume} {132}},\ \bibinfo
  {pages} {1350} (\bibinfo {year} {2013})}\BibitemShut {NoStop}%
\bibitem [{\citenamefont {Bokdam}\ \emph {et~al.}(2014)\citenamefont {Bokdam},
  \citenamefont {Brocks}, \citenamefont {Katsnelson},\ and\ \citenamefont
  {Kelly}}]{Bokdam:prb14b}%
  \BibitemOpen
  \bibfield  {author} {\bibinfo {author} {\bibfnamefont {M.}~\bibnamefont
  {Bokdam}}, \bibinfo {author} {\bibfnamefont {G.}~\bibnamefont {Brocks}},
  \bibinfo {author} {\bibfnamefont {M.~I.}\ \bibnamefont {Katsnelson}}, \ and\
  \bibinfo {author} {\bibfnamefont {P.~J.}\ \bibnamefont {Kelly}},\ }\href
  {\doibase 10.1103/PhysRevB.90.085415} {\bibfield  {journal} {\bibinfo
  {journal} {Phys. Rev. B}\ }\textbf {\bibinfo {volume} {90}},\ \bibinfo
  {pages} {085415} (\bibinfo {year} {2014})}\BibitemShut {NoStop}%
\bibitem [{\citenamefont {Kresse}\ and\ \citenamefont
  {Hafner}(1993)}]{Kresse:prb93}%
  \BibitemOpen
  \bibfield  {author} {\bibinfo {author} {\bibfnamefont {G.}~\bibnamefont
  {Kresse}}\ and\ \bibinfo {author} {\bibfnamefont {J.}~\bibnamefont
  {Hafner}},\ }\href {\doibase 10.1103/PhysRevB.47.558} {\bibfield  {journal}
  {\bibinfo  {journal} {Phys. Rev. B}\ }\textbf {\bibinfo {volume} {47}},\
  \bibinfo {pages} {558} (\bibinfo {year} {1993})}\BibitemShut {NoStop}%
\bibitem [{\citenamefont {Kresse}\ and\ \citenamefont
  {Furthm{\"{u}}ller}(1996)}]{Kresse:prb96}%
  \BibitemOpen
  \bibfield  {author} {\bibinfo {author} {\bibfnamefont {G.}~\bibnamefont
  {Kresse}}\ and\ \bibinfo {author} {\bibfnamefont {J.}~\bibnamefont
  {Furthm{\"{u}}ller}},\ }\href {\doibase 10.1103/PhysRevB.54.11169} {\bibfield
   {journal} {\bibinfo  {journal} {Phys. Rev. B}\ }\textbf {\bibinfo {volume}
  {54}},\ \bibinfo {pages} {11169} (\bibinfo {year} {1996})}\BibitemShut
  {NoStop}%
\bibitem [{\citenamefont {Kresse}\ and\ \citenamefont
  {Joubert}(1999)}]{Kresse:prb99}%
  \BibitemOpen
  \bibfield  {author} {\bibinfo {author} {\bibfnamefont {G.}~\bibnamefont
  {Kresse}}\ and\ \bibinfo {author} {\bibfnamefont {D.}~\bibnamefont
  {Joubert}},\ }\href {\doibase 10.1103/PhysRevB.59.1758} {\bibfield  {journal}
  {\bibinfo  {journal} {Phys. Rev. B}\ }\textbf {\bibinfo {volume} {59}},\
  \bibinfo {pages} {1758} (\bibinfo {year} {1999})}\BibitemShut {NoStop}%
\bibitem [{\citenamefont {Perdew}\ and\ \citenamefont
  {Zunger}(1981)}]{Perdew:prb81}%
  \BibitemOpen
  \bibfield  {author} {\bibinfo {author} {\bibfnamefont {J.~P.}\ \bibnamefont
  {Perdew}}\ and\ \bibinfo {author} {\bibfnamefont {A.}~\bibnamefont
  {Zunger}},\ }\href {\doibase 10.1103/PhysRevB.23.5048} {\bibfield  {journal}
  {\bibinfo  {journal} {Phys. Rev. B}\ }\textbf {\bibinfo {volume} {23}},\
  \bibinfo {pages} {5048} (\bibinfo {year} {1981})}\BibitemShut {NoStop}%
\bibitem [{\citenamefont {Perdew}\ \emph {et~al.}(1996)\citenamefont {Perdew},
  \citenamefont {Burke},\ and\ \citenamefont {Ernzerhof}}]{Perdew:prl96}%
  \BibitemOpen
  \bibfield  {author} {\bibinfo {author} {\bibfnamefont {J.~P.}\ \bibnamefont
  {Perdew}}, \bibinfo {author} {\bibfnamefont {K.}~\bibnamefont {Burke}}, \
  and\ \bibinfo {author} {\bibfnamefont {M.}~\bibnamefont {Ernzerhof}},\ }\href
  {\doibase 10.1103/PhysRevLett.77.3865} {\bibfield  {journal} {\bibinfo
  {journal} {Phys. Rev. Lett.}\ }\textbf {\bibinfo {volume} {77}},\ \bibinfo
  {pages} {3865} (\bibinfo {year} {1996})}\BibitemShut {NoStop}%
\bibitem [{\citenamefont {Dion}\ \emph {et~al.}(2004)\citenamefont {Dion},
  \citenamefont {Rydberg}, \citenamefont {Schr{\"{o}}der}, \citenamefont
  {Langreth},\ and\ \citenamefont {Lundqvist}}]{Dion:prl04}%
  \BibitemOpen
  \bibfield  {author} {\bibinfo {author} {\bibfnamefont {M.}~\bibnamefont
  {Dion}}, \bibinfo {author} {\bibfnamefont {H.}~\bibnamefont {Rydberg}},
  \bibinfo {author} {\bibfnamefont {E.}~\bibnamefont {Schr{\"{o}}der}},
  \bibinfo {author} {\bibfnamefont {D.~C.}\ \bibnamefont {Langreth}}, \ and\
  \bibinfo {author} {\bibfnamefont {B.~I.}\ \bibnamefont {Lundqvist}},\ }\href
  {\doibase 10.1103/PhysRevLett.92.246401} {\bibfield  {journal} {\bibinfo
  {journal} {Phys. Rev. Lett.}\ }\textbf {\bibinfo {volume} {92}},\ \bibinfo
  {pages} {246401} (\bibinfo {year} {2004})}\BibitemShut {NoStop}%
\bibitem [{\citenamefont {Thonhauser}\ \emph {et~al.}(2007)\citenamefont
  {Thonhauser}, \citenamefont {Cooper}, \citenamefont {Li}, \citenamefont
  {Puzder}, \citenamefont {Hyldgaard},\ and\ \citenamefont
  {Langreth}}]{Thonhauser:prb07}%
  \BibitemOpen
  \bibfield  {author} {\bibinfo {author} {\bibfnamefont {T.}~\bibnamefont
  {Thonhauser}}, \bibinfo {author} {\bibfnamefont {V.~R.}\ \bibnamefont
  {Cooper}}, \bibinfo {author} {\bibfnamefont {S.}~\bibnamefont {Li}}, \bibinfo
  {author} {\bibfnamefont {A.}~\bibnamefont {Puzder}}, \bibinfo {author}
  {\bibfnamefont {P.}~\bibnamefont {Hyldgaard}}, \ and\ \bibinfo {author}
  {\bibfnamefont {D.~C.}\ \bibnamefont {Langreth}},\ }\href {\doibase
  10.1103/PhysRevB.76.125112} {\bibfield  {journal} {\bibinfo  {journal} {Phys.
  Rev. B}\ }\textbf {\bibinfo {volume} {76}},\ \bibinfo {pages} {125112}
  (\bibinfo {year} {2007})}\BibitemShut {NoStop}%
\bibitem [{\citenamefont {Klime{\v{s}}}\ \emph {et~al.}(2011)\citenamefont
  {Klime{\v{s}}}, \citenamefont {Bowler},\ and\ \citenamefont
  {Michaelides}}]{Klimes:prb11}%
  \BibitemOpen
  \bibfield  {author} {\bibinfo {author} {\bibfnamefont {J.}~\bibnamefont
  {Klime{\v{s}}}}, \bibinfo {author} {\bibfnamefont {D.~R.}\ \bibnamefont
  {Bowler}}, \ and\ \bibinfo {author} {\bibfnamefont {A.}~\bibnamefont
  {Michaelides}},\ }\href {\doibase 10.1103/PhysRevB.83.195131} {\bibfield
  {journal} {\bibinfo  {journal} {Phys. Rev. B}\ }\textbf {\bibinfo {volume}
  {83}},\ \bibinfo {pages} {195131} (\bibinfo {year} {2011})}\BibitemShut
  {NoStop}%
\bibitem [{\citenamefont {Ruiz}\ \emph {et~al.}(2012)\citenamefont {Ruiz},
  \citenamefont {Liu}, \citenamefont {Zojer}, \citenamefont {Scheffler},\ and\
  \citenamefont {Tkatchenko}}]{Ruiz:prl12}%
  \BibitemOpen
  \bibfield  {author} {\bibinfo {author} {\bibfnamefont {V.~G.}\ \bibnamefont
  {Ruiz}}, \bibinfo {author} {\bibfnamefont {W.}~\bibnamefont {Liu}}, \bibinfo
  {author} {\bibfnamefont {E.}~\bibnamefont {Zojer}}, \bibinfo {author}
  {\bibfnamefont {M.}~\bibnamefont {Scheffler}}, \ and\ \bibinfo {author}
  {\bibfnamefont {A.}~\bibnamefont {Tkatchenko}},\ }\href {\doibase
  10.1103/PhysRevLett.108.146103} {\bibfield  {journal} {\bibinfo  {journal}
  {Phys. Rev. Lett.}\ }\textbf {\bibinfo {volume} {108}},\ \bibinfo {pages}
  {146103} (\bibinfo {year} {2012})}\BibitemShut {NoStop}%
\bibitem [{\citenamefont {Bj{\"{o}}rkman}\ \emph {et~al.}(2012)\citenamefont
  {Bj{\"{o}}rkman}, \citenamefont {Gulans}, \citenamefont {Krasheninnikov},\
  and\ \citenamefont {Nieminen}}]{Bjorkman:prl12}%
  \BibitemOpen
  \bibfield  {author} {\bibinfo {author} {\bibfnamefont {T.}~\bibnamefont
  {Bj{\"{o}}rkman}}, \bibinfo {author} {\bibfnamefont {A.}~\bibnamefont
  {Gulans}}, \bibinfo {author} {\bibfnamefont {A.~V.}\ \bibnamefont
  {Krasheninnikov}}, \ and\ \bibinfo {author} {\bibfnamefont {R.~M.}\
  \bibnamefont {Nieminen}},\ }\href {\doibase 10.1103/PhysRevLett.108.235502}
  {\bibfield  {journal} {\bibinfo  {journal} {Phys. Rev. Lett.}\ }\textbf
  {\bibinfo {volume} {108}},\ \bibinfo {pages} {235502} (\bibinfo {year}
  {2012})}\BibitemShut {NoStop}%
\bibitem [{\citenamefont {Mittendorfer}\ \emph {et~al.}(2011)\citenamefont
  {Mittendorfer}, \citenamefont {Garhofer}, \citenamefont {Redinger},
  \citenamefont {Klime{\v{s}}}, \citenamefont {Harl},\ and\ \citenamefont
  {Kresse}}]{Mittendorfer:prb11}%
  \BibitemOpen
  \bibfield  {author} {\bibinfo {author} {\bibfnamefont {F.}~\bibnamefont
  {Mittendorfer}}, \bibinfo {author} {\bibfnamefont {A.}~\bibnamefont
  {Garhofer}}, \bibinfo {author} {\bibfnamefont {J.}~\bibnamefont {Redinger}},
  \bibinfo {author} {\bibfnamefont {J.}~\bibnamefont {Klime{\v{s}}}}, \bibinfo
  {author} {\bibfnamefont {J.}~\bibnamefont {Harl}}, \ and\ \bibinfo {author}
  {\bibfnamefont {G.}~\bibnamefont {Kresse}},\ }\href {\doibase
  10.1103/PhysRevB.84.201401} {\bibfield  {journal} {\bibinfo  {journal} {Phys.
  Rev. B}\ }\textbf {\bibinfo {volume} {84}},\ \bibinfo {pages} {201401}
  (\bibinfo {year} {2011})}\BibitemShut {NoStop}%
\bibitem [{\citenamefont {Reguzzoni}\ \emph {et~al.}(2012)\citenamefont
  {Reguzzoni}, \citenamefont {Fasolino}, \citenamefont {Molinari},\ and\
  \citenamefont {Righi}}]{Reguzzoni:prb12}%
  \BibitemOpen
  \bibfield  {author} {\bibinfo {author} {\bibfnamefont {M.}~\bibnamefont
  {Reguzzoni}}, \bibinfo {author} {\bibfnamefont {A.}~\bibnamefont {Fasolino}},
  \bibinfo {author} {\bibfnamefont {E.}~\bibnamefont {Molinari}}, \ and\
  \bibinfo {author} {\bibfnamefont {M.~C.}\ \bibnamefont {Righi}},\ }\href
  {\doibase 10.1103/PhysRevB.86.245434} {\bibfield  {journal} {\bibinfo
  {journal} {Phys. Rev. B}\ }\textbf {\bibinfo {volume} {86}},\ \bibinfo
  {pages} {245434} (\bibinfo {year} {2012})}\BibitemShut {NoStop}%
\bibitem [{\citenamefont {Berland}\ and\ \citenamefont
  {Hyldgaard}(2013)}]{Berland:prb13}%
  \BibitemOpen
  \bibfield  {author} {\bibinfo {author} {\bibfnamefont {K.}~\bibnamefont
  {Berland}}\ and\ \bibinfo {author} {\bibfnamefont {P.}~\bibnamefont
  {Hyldgaard}},\ }\href {\doibase 10.1103/PhysRevB.87.205421} {\bibfield
  {journal} {\bibinfo  {journal} {Phys. Rev. B}\ }\textbf {\bibinfo {volume}
  {87}},\ \bibinfo {pages} {205421} (\bibinfo {year} {2013})}\BibitemShut
  {NoStop}%
\bibitem [{Note1()}]{Note1}%
  \BibitemOpen
  \bibinfo {note} {Experiments concur that the interaction is weak but the
  binding separation has not been measured directly \cite
  {Preobrajenski:ss05,Joshi:nanol12,Roth:nanol13}; the value reported in
  Ref.~\cite {Joshi:nanol12} results from interpreting STM measurements with
  the help of DFT calculations.}\BibitemShut {Stop}%
\bibitem [{\citenamefont {Roth}\ \emph {et~al.}(2013)\citenamefont {Roth},
  \citenamefont {Matsui}, \citenamefont {Greber},\ and\ \citenamefont
  {Osterwalder}}]{Roth:nanol13}%
  \BibitemOpen
  \bibfield  {author} {\bibinfo {author} {\bibfnamefont {S.}~\bibnamefont
  {Roth}}, \bibinfo {author} {\bibfnamefont {F.}~\bibnamefont {Matsui}},
  \bibinfo {author} {\bibfnamefont {T.}~\bibnamefont {Greber}}, \ and\ \bibinfo
  {author} {\bibfnamefont {J.}~\bibnamefont {Osterwalder}},\ }\href {\doibase
  10.1021/nl400815w} {\bibfield  {journal} {\bibinfo  {journal} {Nano Letters}\
  }\textbf {\bibinfo {volume} {13}},\ \bibinfo {pages} {2668} (\bibinfo {year}
  {2013})}\BibitemShut {NoStop}%
\end{thebibliography}
\end{document}